\documentclass[a4paper,11pt]{paper}
\usepackage[latin1]{inputenc}
\usepackage[english]{babel}
\usepackage{amsmath,epsfig,dcolumn,times}
\usepackage{calc}
\usepackage{caption}
\usepackage{amsmath}
\usepackage{amsfonts}
\usepackage{amssymb}
\usepackage{epsfig}
\usepackage{calc}
\usepackage{subfigure}
\usepackage{array}
\usepackage{longtable}
\usepackage{moreverb}
\usepackage{tabularx}

\newcommand{\bbm}{\begin{bmatrix}}
\newcommand{\ebm}{\end{bmatrix}}

\renewcommand{\comment}[1]{}


{\begin{list}{}{%
      \settowidth{\labelwidth}{\textsf{#1}xxx}%
      \setlength{\leftmargin}{\labelwidth+\labelsep}}}%
  {\end{list}}

\newcommand{\ul}[1]{\underline{#1}}

\begin{document}

\twocolumn[

	\begin{center} 	
	\bf
		HIGH PRECISION MODELING TOWARDS THE $10^{-20}$ LEVEL \\
		\vspace{0.5cm}
		\vspace{0.5cm}
	{\scriptsize\bf
	Author: M. Andres$^1$, L. Banz$^1$, A. Costea$^1$, E. Hackmann$^2$, S. Herrmann$^2$, C. L\"ammerzahl$^2$, L. Nesemann$^1$, B. Rievers$^2$, E. P. Stephan$^1$ \\
	1 : Institute for Applied Mathematics, Leibniz University Hannover, Germany\\
	2 : Center of Applied Space Technology and Microgravity ZARM, University of Bremen, Germany

\vspace*{0.2cm}

	}
	
\vspace*{0.3cm}

\end{center}
\vspace*{-0.3cm}
\begin{abstract}
The requirements for accurate numerical simulation are increasing constantly. Modern high precision physics experiments now exceed the achievable numerical accuracy of standard commercial and scientific simulation tools. One example are optical resonators for which changes in the optical length are now commonly measured to $10^{-15}$ precision \cite{Schiller09}. The achievable measurement accuracy for resonators and cavities is directly influenced by changes in the distances between the optical components. If deformations in the range of $10^{-15}$ occur, those effects cannot be modeled and analysed any more with standard methods based on double precision data types. New experimental approaches point out that the achievable experimental accuracies may improve down to the level of $10^{-17}$ in the near future \cite{Brueckneretal10}. For the development and improvement of high precision resonators and the analysis of experimental data, new methods have to be developed which enable the needed level of simulation accuracy. Therefore we plan the development of new high precision algorithms for the simulation and modeling of thermo-mechanical effects with an achievable accuracy of $10^{-20}$. In this paper we analyse a test case and identify the problems on the way to this goal. 
\end{abstract}\vspace*{0.3cm}
]

\comment{
\section*{\ul{Nomenclature}}
\begin{center}
	\begin{tabular}{l l l l}
		& FE & Finite Element \\
		&  &  &
	\end{tabular}

\end{center}

}

\section*{\ul{Motivation}}

Optical high-finesse resonators are widely used as a frequency reference for the stabilization of lasers e.g. in optical atomic clocks, now aiming for the $10^{-18}$ level of precision \cite{Rosenband}, or in direct tests of special and general relativity. For example, \cite{Schiller09} and \cite{Herrmann09} tested the isotropy of the speed of light by comparing the frequencies of two lasers referenced to crossed resonators with a relative precision of $10^{-15}$ at 1\,s integration time.

Typically, an electronic feedback loop is applied to actively stabilize the laser frequency to the optical resonator. The laser frequency then results directly from the geometrical resonator dimensions as given by $\nu_L =mc=2L$, where $L$ is the resonator length and $m$  is the longitudinal mode number. The measurement performance thus depends on the length stability of the implemented cavities. Since $\Delta \nu/\nu = \Delta L/L$, high precision frequency stabilization at the $\,10^{-15}\;$ level and below, requires a relative cavity length stability of the same magnitude. The dominant effects which cause expansions or contractions are gravity (in particular if an experiment is assembled on earth and conducted in space) and thermal expansion of the components. Therefore high precision modeling of thermo-mechanical effects is a basic requirement for a safe design and verification of optical resonators. Only if the models and simulations offer the same precision which can be achieved in experiments, the optimization, design, and evaluation of optical resonators can be performed in a sound way.

To date, the best cavity stabilized lasers have achieved a relative frequency stability on the order of few parts in $10^{16}$ \cite{Young}. As pointed out by Numata et al.~\cite{Numata,Notcutt}, this precision is limited by fundamental thermal noise, mostly affecting the highly reflective mirror coatings. Consequently, low-noise, single-crystalline silicon mirrors are currently being developed \cite{Brueckneretal10} and an improvement of optical cavity stability down to the $10^{-17}$ level might become feasible already within the next years. These cavities will ultimately help to build optical clocks of improved accuracy or to carry out fundamental tests at the $10^{-17}$ level and below. This development however needs to be paralleled by similar\, advancements\, in\, high\, precision\, modeling tools.

In order to react to the grown achievable accuracies of optical resonators and the need for improved modeling methods, we intend to develop a new method for the modeling of thermo-mechanical effects in 3 dimensions with an achievable numerical accuracy of $10^{-20}$. In this paper, we outline the difficulties to overcome by analysing a cavity test case. 

\section*{\ul{Test case configuration}}
In the following we will present the cavity test case which will be used to demonstrate the problems for modeling of thermo-mechanical effects with a numerical accuracy of $10^{-20}$. The analysis is confined to a single mirror within the cavity and performed using two different approaches.

\begin{figure*}
\centering
\includegraphics[width=0.9\textwidth]{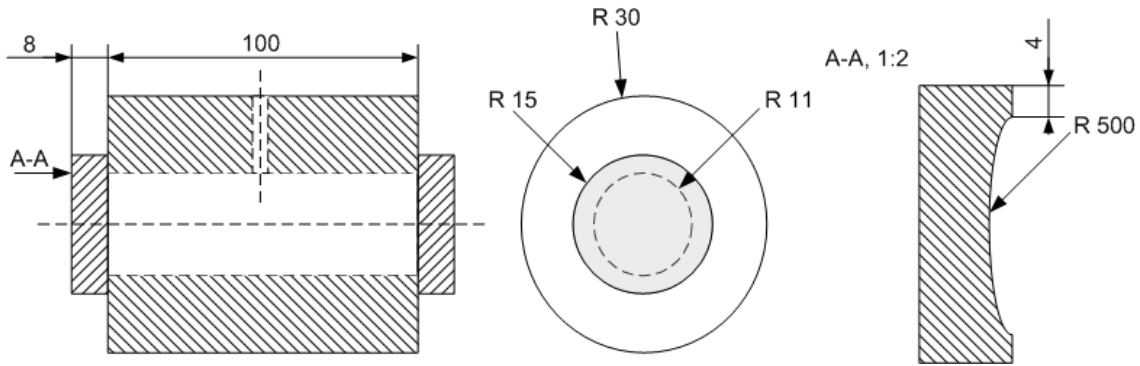}
\caption{Design of cavity test case configuration.}
\label{designTC}
\end{figure*}

\subsection*{Cavity test case}
An optical cavity is typically axially symmetric and composed of two highly reflective\, concave\, lenses\, which are bonded to a spacer. The dimensions of the test cavity analyzed here are adapted from Webster et al.~\cite{Websteretal07}, for the design see Figure ~\ref{designTC}. The mirror substrates, the mirror coatings, and the spacer are assumed to be made of ultra-low expansion glass (ULE) provided by Corning with the relevant material parameters listed in Tab.~\ref{material}. Although ULE is known for anisotropic thermal expansion, we assume as an approximation that all materials are homogenous and isotropic and that the space around the cavity is a vacuum at room temperature. Further, we take only thermo-elastic effects into account and neglect other error sources like vibration or material losses.

\begin{table}[th]
\begin{tabular}{p{0.3\textwidth}|c}
Material parameter & value\\
\hline
\hline
Thermal conductivity [W/mK] & $1.31$\\
\hline
Specific heat [J/kgK] & $767$\\
\hline
Mean coefficient of thermal expansion [1/K] & $0 \pm 30  \times 10^{-9}$\\
\hline
Elastic modulus [Pa] & $67.6 \times 10^9$\\
\hline
Poisson's ratio & $0.17$\\
\hline
Mass density $\rm [kg/m^3]$ & $2.21 \times 10^3$\\
\hline
\end{tabular}
\caption{Material parameters of ULE Corning code 7972.}
\label{material}
\end{table}

\subsection*{Modeled effects, boundaries, and loads}
Regarding the setup of the test cavity, the relevant thermal physical effects are heat conduction and radiation as well as the thermal expansion of the material. Due to the assumed room temperature, heat radiation may be neglected as a first simplification. The spacer may radiate to the whole surrounding space and, thus, we assume it as a heat sink with a constant temperature of $300 K$ as a second simplification. Third, we assume that the spacer is fixed in space and cannot be deformed. As the mirrors are bonded to the spacer, a deformation of the mirrors at the contact areas is also not allowed. However, heat conduction at the contact area is taken into account. 

ULE is designed to have a minimum thermal expansion coefficient at some temperature $T_{\rm ULE}$ near room temperature, whose exact value has to be determined experimentally for each cavity. Even if $T_{\rm ULE}$ is exactly known, in practice it is not possible to achieve a stable temperature of $T_{\rm ULE}$ over the whole cavity, and a small temperature gradient is inevitable. However, by careful thermal control a temperature stability of a few $\rm mK$ can be obtained if operating near $T_{\rm ULE}$ and even about $100 \rm \mu K$ for temperatures around $300 \rm K$ \cite{Alnisetal08}. Here we assume the mirrors substrates and coatings to have a stable temperature corresponding to a coefficient of thermal expansion of $30 \times 10^{-9} \rm K^{-1}$. For more rigorous treatments than this test case the thermal expansion coefficient should be varied in time to reflect the small disturbances of the temperature.

Also, we assume that this test cavity has a finesse of $F=10^6$ and a transmission coefficient of the incoupling mirror of $T = 10^{-6}$. For an input power of $P_{\rm in} = 50 \mu W$ this results with the gain factor of
\begin{equation}
\frac{F^2}{\pi^2} T \approx 10^5
\end{equation}
in an effective power of about $P = 5 W$ stored in the cavity. The losses $A$ due to absorption corresponding to the chosen values for $F \approx \frac{\pi}{T+A}$ and $T$ can be assumed to $A \approx 10^{-6}$ resulting in a dissipated power of about $5 \mu W$ per mirror. The laser is assumed here to have a diameter of $200 \mu m$ and to hit the mirror perfectly at its center.

Once an equilibrium state is reached, the temperature is constant in time but may vary over the cavity. However, due to the changed temperature the cavity will deform, what in turn changes the boundary conditions for the heat conduction. Therefore, the final static state has to be determined iteratively. For this test case, we neglect the influence of the changes of boundary conditions and perform one static analysis, only.

\section*{\ul{FE models and results}}
The test case described above was modeled using the Finite Element (FE) approach. Two independ programs were used for this task: the academic version of the commercial ANSYS software and the software package MaiProgs developed at the IfAM. 

\subsection*{ANSYS model}
\begin{figure}
\centering
\includegraphics[width=0.4\textwidth]{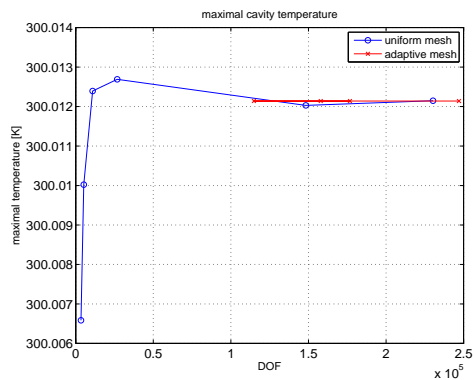}
\caption{Max. temp vs. DOF using ANSYS}\label{ANSYS_maxtemp}
\end{figure}

\begin{figure}
\centering
\includegraphics[width=0.4\textwidth]{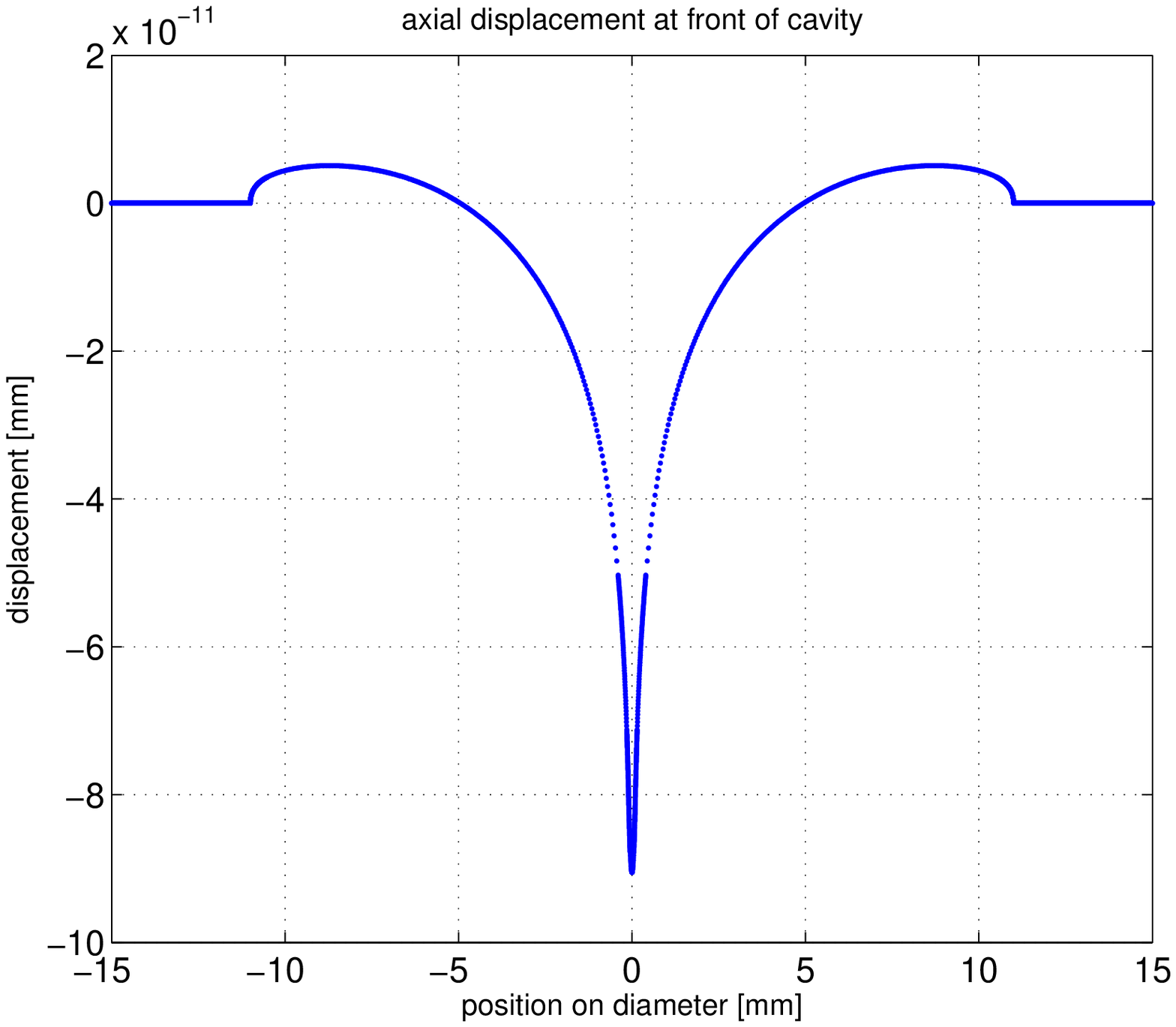}
\caption{Nodal solution plot using ANSYS}\label{ANSYS_dis1}
\end{figure}

\begin{figure}
\centering
\includegraphics[width=0.4\textwidth]{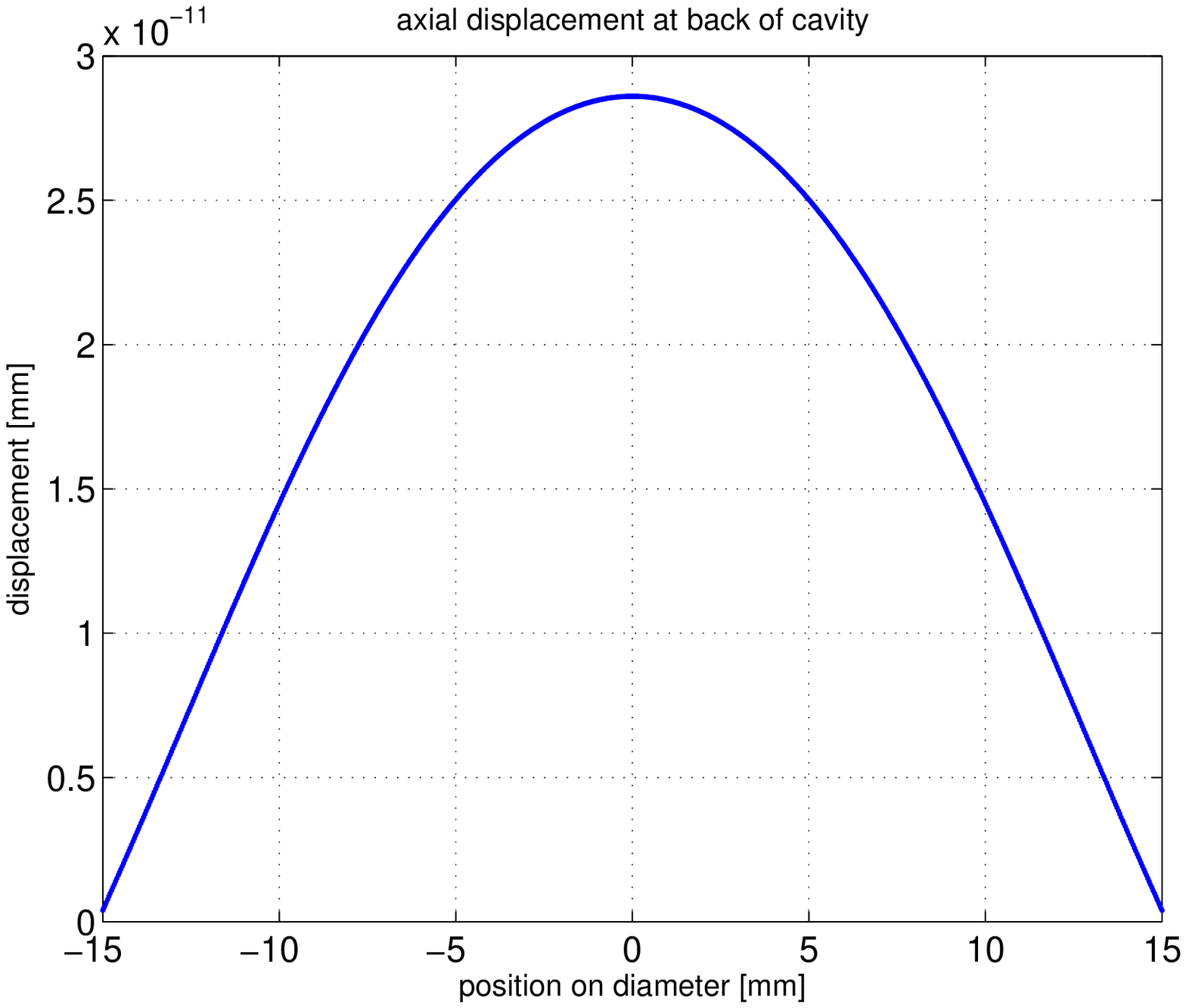}
\caption{Nodal solution plot using ANSYS}\label{ANSYS_dis2}
\end{figure}

As a first approach, one half of the axially symmetric cavity was modeled in three dimensions. However, due to the limited node number of $256000$ nodes of the academic version of ANSYS, this turned out to be nonsufficient for reaching a high modeling accuracy. Therefore, the axial symmetry of the problem was used to reduce the model to two dimensions. For the computation of heat conduction and mechanical deformation, the eight nodes PLANE77 and PLANE183 element types were used. On the spacer a fixed temperature and no displacement are prescribed and, therefore, we modeled it in ANSYS with a uniform mesh with a large maximal element size of $1 \rm mm$. On the lens we started with the same element size and reduced it step by step. However, the best results could be obtained by using a mesh adapted to the problem with maximal element sizes of approximatly $0.002 \rm mm$ in the regions of interest and a total node number of about $247000$. The maximum temperatures for the various degrees of freedom are shown in figure \ref{ANSYS_maxtemp}. After solving the thermal problem the resulting temperature distribution was applied as a body load for the following structural analysis. From a preliminary analysis the displacements were expected in the range of $10^{-12} \rm mm$. Therefore, as the structural analysis is a linear problem, we decided to rescale the thermal expansion coefficient for a better numerical performance. The resulting displacements at front and back of the cavity are shown in figures \ref{ANSYS_dis1} and \ref{ANSYS_dis2}.



\subsection*{MaiProgs model}
\begin{figure}
\centering
\includegraphics[width=0.4\textwidth]{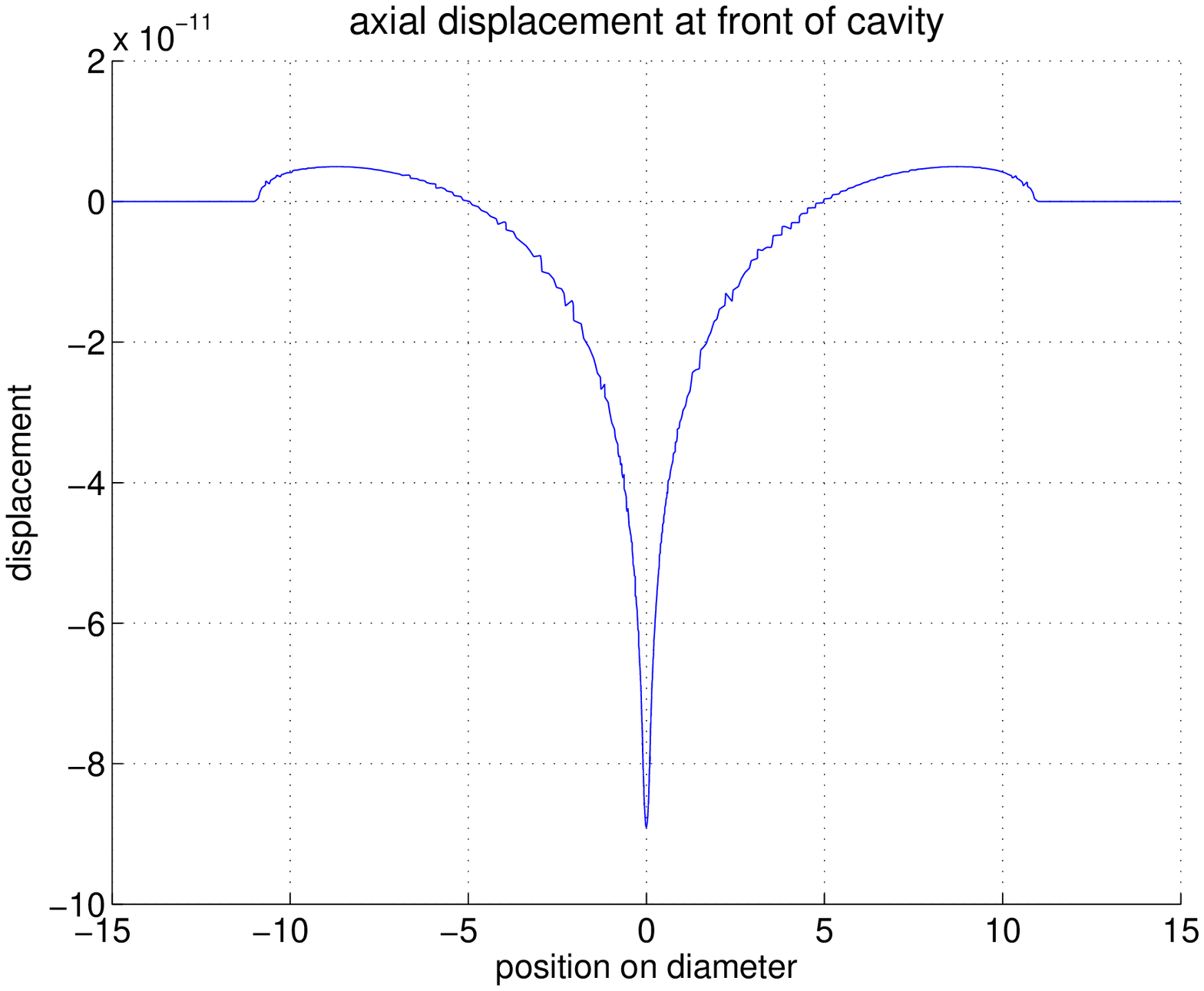}
\caption{Nodal solution plot using MaiProgs}\label{MaiProgs_dis1}
\end{figure}

\begin{figure}
\centering
\includegraphics[width=0.4\textwidth]{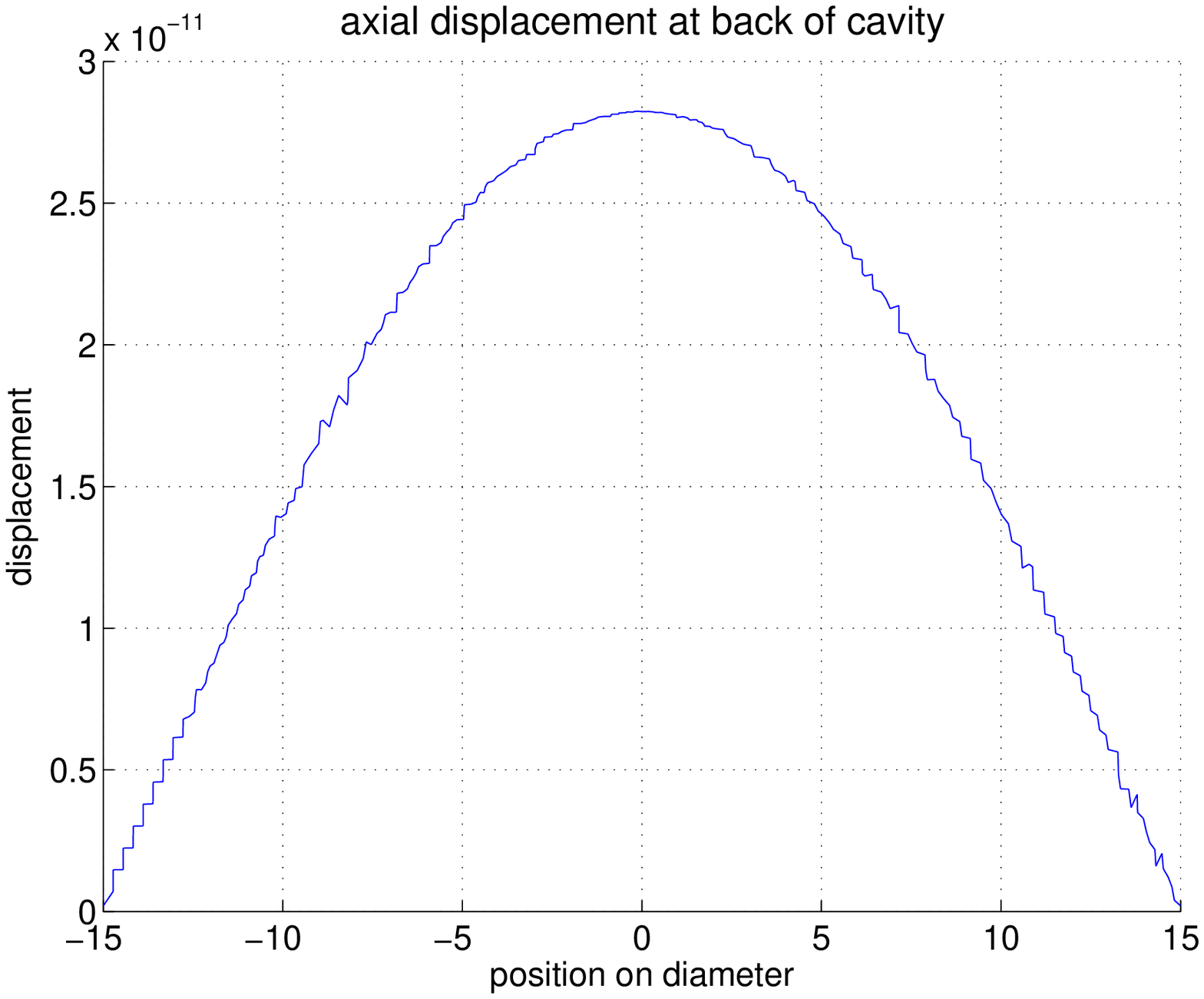}
\caption{Nodal solution plot using MaiProgs}\label{MaiProgs_dis2}
\end{figure}

At the IfAM computations were made for linear elements on tetrahedral meshes with up to seven million nodes. For the computations the self developed FE/BE-software MaiProgs \cite{MaiProgs} was used. In Figures \ref{MaiProgs_dis1} and \ref{MaiProgs_dis2} the axial displacements on the cross section of the cavity at the front and the back is plotted using the data of the computations with MaiProgs. 



\begin{figure}
\centering
\includegraphics[width=0.4\textwidth]{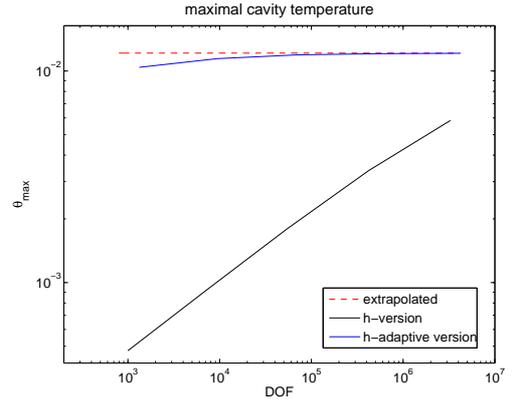}
\caption{Max. temp vs. element size}\label{MaiProgs_maxtemp}
\end{figure}

To point out the sensitivity of the mathematical model on the accuracy of the solution, computations on two different meshes were performed at the IfAM. In the first case the usual h-version on a quasiuniform mesh was computed. The geometry is modeled via two cylinders where the ring of the outer cylinder fits with the ring of the cavity that is bonded to the spacer. The tetrahedral elements have nearly the same shape and size. Since the laser is striking the cavitiy in its midpoint with a radius of $0.1 \rm mm$ it is obvious that the temperature will have a large gradient there. This is also emphasized by the axial displacement in Figure \ref{MaiProgs_dis1} where the displacement seems to have a singularity in the midpoint of the cavity. Using this information the IfAM modeled a second mesh consisting of three cylinders where the inner cylinder has the radius of the laser such that the coarsest mesh has 400 more points in the neighborhood of the midpoint than the first version. The tetrahedral elements do not have the same size any longer, since the elements near the midpoint of the cavity are considerably smaller than the other elements. From this point of view the second mesh can be regarded as kind of an adaptive mesh. This leads to a better approximation of the boundary loads and hence to a better approximation of the solution as can be seen in Figure \ref{MaiProgs_maxtemp}. Here the red dashed line is the extrapolated maximal temperature inside the cavity. On both meshes the IfAM computed four steps of the h-version \cite{Stephan96, StephanVL} starting with a coarsest mesh with 2400 points and 2800 points, respectively. Figure \ref{MaiProgs_maxtemp} shows that the computations on the second mesh lead to much better approximations of the temperature. It seems reasonable to assume that an adaptive computation using some kind of a posteriori error estimator (residual, hierarchical, etc.) \cite{Stephan04,Arayaetal05} would exceedingly improve the accuracy of the solution.

\subsection*{Comparison of results}
The two different softwares and approaches described in this section lead to results comparing well to each other. However, a closer inspection of the two results reveal significant differences: the maximal temperatures for the two results differ by $$\Delta_T = 1.904041846501059 \times 10^{-05} \, \rm K \,.$$ Given this quite large discrepancy it is almost suprising that the maximal displacements for both approaches differ by only $$\Delta_D=1.447725122807002 \times 10^{-12} \, \rm mm \,.$$ These differences are caused on the one hand by the discretization errors connected to the different meshes, and on the other hand by the limited accuracy of the numerical procedures involved in the solution process. Whereas the first error source can be minimized by using finer meshes, the second one is an inert problem of floating point arithmetic. As a rule of thumb, one can say that for a result with $n$ significant accurate digits an accuracy of $2n$ significant digits for each addition and multiplication is needed. As the exponent is treated seperatly, for linear problems with known order of magnitude for displacement or temperature the accurate digits can be considerably improved by a suitable rescaling of the material properties. However, for nonlinear problems this approach will in general not work. As a consequence, the only way to minimize the numerical errors is to use quad (with 128 bits) or even  arbitrary precision arithmetic instead of the usual double precision (with 64 bits). Since there is no appropriate hardware available, these computations must be performed with a software emulation. This slows down the computation process significantly.

\section*{\ul{Accuracy Evaluation}}

\subsection*{ANSYS}
\begin{center}
\begin{figure}
\centering
\includegraphics[width=0.4\textwidth]{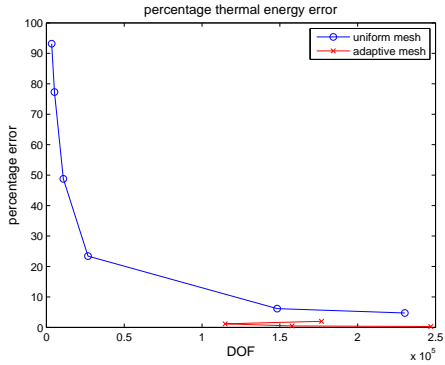}
\caption{Thermal energy error norm plot}
\label{ANSYS_therr}
\end{figure}
\end{center}

\begin{center}
\begin{figure}
\centering
\includegraphics[width=0.4\textwidth]{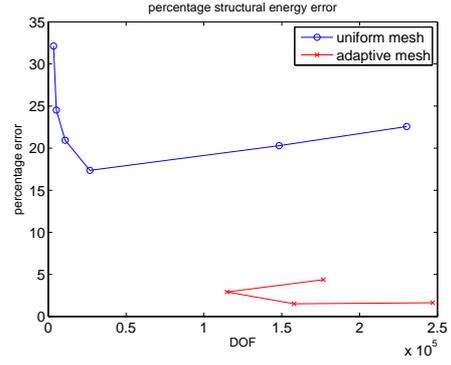}
\caption{Structural energy error norm plot}
\label{ANSYS_sterr}
\end{figure}
\end{center}

Within the ANSYS software a posteriori error estimator can be used to evaluate the quality of the solution. As the exact solution is not known in general and, in particular, not for our test case, this error estimator compares the (discountinous) heat flux or stresses, respectively, in each node as computed by ANSYS to an averaged (continuous) function, which interpolates these values. In each element $E_i$ the energy norm $e_i$ of the difference of the discontinuous values and the interpolated function is a measure for the error in this element. For the whole model a percentaged relative energy norm error can be given by the sum of all $e_i$'s divided by $\sum_i e_i$ and the thermal dissipation energy. This is a kind of Zienkewics/Zhu error estimator \cite{ZienkewicsZhu87}, which converges to the true error. The percentage error for various degree of freedoms are plotted in Figures \ref{ANSYS_therr} and \ref{ANSYS_sterr} for both the thermal and structural computations.

\subsection*{MaiProgs}
\begin{center}
\begin{figure}
\centering
\includegraphics[width=7.0cm,angle = 0]{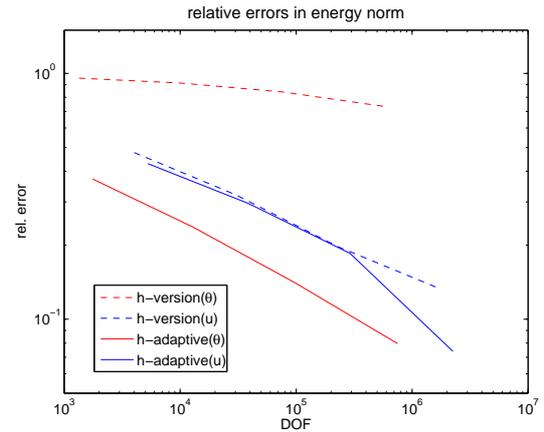}
\caption{Error norm plot}\label{MaiProgs_error}
\end{figure}
\end{center}

Since the exact solution of the presented test case is not known in advance it is impossible to compute the exact error of the computed approximations in the energy norm. Therefore Figure in \ref{MaiProgs_error} extrapolated norms for the temperature and the displacement are compared with the energy norms of the approximated solutions. The error does not reflect the accuracy of the solution but can be considered as a measure to compare different approximations. The aim is to derive efficient and reliable a posteriori error estimators for the test case in order to have a better prediction of the true error. Of course, these estimators can be used within adaptive algorithms leading to a faster convergence \cite{Stephan04,Arayaetal05}.

\section*{\ul{Conclusion}}
We analysed a cavity test case for thermal induced changes in the cavity length, which directly influences the stability of the laser frequency used to interrogate the cavity. We used two different and independent softwares for a finite element analysis of the thermo-elastic deformation of the cavity lens due to the dissipated power of the laser. 

Despite the numerous simplifications assumed for this test case the results of the two independently computed values for the maximal displacement differ by an order of magnitude of $10^{-12}$, which is considerably lower than the corresponding achievable frequency stabilities of ultrastable cavities. The individual analysis have two major error sources: the discretization error of the mesh and the numerical error of the solution process. The first error can be shown to converge to zero for an infinitly fine mesh. Successively finer meshes can be produced with any commercial FE software and adequate computing power. Therefore, this error source can be minimized without principle problems. The second error source is inert to floating point arithmetic and can only be minimized if data types with more bits are used. For the present hardware, these data types have in general to be software emulated, which will slow down the computation process considerably. Also, to the knowledge of the authors, a publicly available FE-solver using quad or arbitrary precision arithmetics does not exist.

The test case presented here is processed taking into account only thermo-elastic effects. For the design of high precision cavities, this is of course not sufficient. However, the principle problems for the FE-modelling will remain the same. Therefore, cavities with length stabilities on the $10^{-17}$ level or even higher can only be simulated if the problem of numerical errors can be solved. We plan to solve this problem with the development of a stand-alone FE-solver using arbitrary precision arithmetics, which is based on the solver already implemented in MaiProgs, together with an interface to ANSYS or other prepocessors.



\begin{thebibliography}{}
\bibitem{Schiller09} 
Ch. Eisele, A. Yu. Nevsky, and S. Schiller. {\it{Laboratory Test of the Isotropy of Light Propagation at the $10^{-17}$ Level}}, Phys. Rev. Letters 103, 090401 (2009).

\bibitem{Brueckneretal10}
F. Br\"uckner, D. Friedrich, T. Clausnitzer, M. Britzger, O. Burmeister, K. Danzmann, E-B. Kley, A. T\"unnermann, and R. Schnabel. {\it Realization of a Monolithic High-Reflectivity Cavity Mirror from a Single Silicon Crystal}, Phys. Rev. Lett. 104, 163903 (2010).

\bibitem{Rosenband} T. Rosenband et al. {\it Frequency Ratio of Al+ and Hg+ Single-Ion Optical Clocks; Metrology at the 17th Decimal Place},
Science 319, 1808 -- 1812 (2008).

\bibitem{Herrmann09}
S. Herrmann, A. Senger, K. M\"ohle, M. Nagel, E.V. Kovalchuk, and A. Peters. {\it{Rotating optical cavity experiment testing Lorentz invariance at the $10-17$ level}}, Phys. Rev. D 80, pp. 105011 (2009).

\bibitem{Young} B. C. Young, F. C. Cruz, W. M. Itano, and J. C. Bergquist. {\it Visible Lasers with Subhertz Linewidths}, Phys. Rev. Lett. 82, 3799-3802.

\bibitem{Numata} K. Numata, A. Kemery, and J. Camp. {\it Thermal-Noise Limit in the Frequency Stabilization of Lasers with Rigid Cavities}, Phys. Rev. Lett. 93, 250602 (2004).

\bibitem{Notcutt} M. Notcutt, L-S. Ma, A. D. Ludlow, S. M. Foreman, J. Ye, and J. L. Hall. {\it Contribution of thermal noise to frequency stability of rigid optical cavity via Hertz-linewidth lasers}, Phys. Rev. A 73, 031804(R) (2006).

\bibitem{Websteretal07} S.A. Webster, M. Oxborrow, and P.Gill. {\it Vibration insensitive optical cavity}, Phys.Rev. A75, 011801, 2007. 

\bibitem{Alnisetal08}
J. Alnis, A. Matveev, N. Kolachevsky, Th. Udem, and T. W. H\"ansch. {\it Subhertz linewidth diode lasers by stabilization to vibrationally and thermally compensated ultralow-expansion glass Fabry-P\'{e}rot cavities}, Phys. Rev. A 77, 053809 (2008).

\bibitem{MaiProgs}
M. Maischak. {\it Webpage of the Software package Maiprogs}, http://www.ifam.uni-hannover.de/~maiprogs.

\bibitem{Stephan96}
E.P. Stephan. {\it The h-p version of the boundary element method for solving 2- and 3-dimensional problems}, Comp. Meth. Appl. Mech. Eng. 133 (1996) 183-208. 

\bibitem{StephanVL}
E.P. Stephan. {\it Hp-Finite Element Methoden}, Vorlesungsskript 2009, Leibniz Universit\"at Hannover.

\bibitem{Stephan04}
E.P. Stephan. {\it Coupling of Boundary Element Methods and Finite Element Methods}, Encyclopedia of Computational Mechanics, Edited by Erwin Stein, René de Borst and Thomas J.R. Hughes. Vol. 1, Chapter 13: Fundamentals. 2004 John Wiley \& Sons.

\bibitem{Arayaetal05}
R. Araya, A.H. Poza, and E.P. Stephan. {\it  A hierarchical a posteriori error estimate for an advection-diffusion-reaction problem}, Math. Models Methods Appl. Sci. 15 (2005), no. 7, 1119-1139. 

\bibitem{ZienkewicsZhu87}
0.C. Zienkiewics and J.Z. Zhu. {\it A simple error estimator and adaptive procedure for practical engineering analysis}, Int. J. Numer. Meth. Eng. 24, 337-357 (1987).


\bibitem{Fox09}
R. W. Fox. {\it Temperature analysis of low-expansion Fabry-Perot cavities}, Optics Express 17, 15023 (2009).



\end{thebibliography}
\end{document}